\documentclass[preprint,aps]{revtex4}

\usepackage{graphicx}
\usepackage{dcolumn}
\usepackage{bm}

\begin{document}

\title{Annihilation Contributions and CP Asymmetries\\
in $B^+ \to \pi^+ K^0, K^+ K^0$ and $B^0\to K^0 \bar K^0$}

\author{Xi-Qing Hao$^1$, Xiao-Gang He$^{1,2}$ and Xue-Qian Li$^1$}
\affiliation{$^1$Department of Physics, Nankai University, Tianjin\\
$^2$Department of Physics and Center for Theoretical Sciences,
National Taiwan University, Taipei}

\date{\today} 

\begin{abstract}
Recently the branching ratios for $B^+\to K^+\bar K^0$ and $B^0
\to K^0 \bar K^0$ have been measured. Data indicate that the
annihilation amplitudes in these decays are not zero. A non-zero
annihilation amplitude plays an important role in CP violation for
$B^+\to \pi^+ K^0, K^+ \bar K^0$. Using the measured branching
ratios for these decays, we show that there is an absolute bound
of $5\%$ for the size of CP asymmetry in $B^+\to \pi^+ K^0$ from a
relation between the amplitudes of these decays. The size of CP
asymmetry in $B^+ \to K^+\bar K^0$ can, however, be as large as
$90\%$. Future experimental data will test these predictions.
\end{abstract}

\maketitle

Rare $B$ decays provide much information about the Standard Model
(SM) of the strong and electroweak interaction. So far the SM is
perfectly consistent with data on $B$ decays. With more data
becoming available from Babar and Belle experiments, $B$ physics
has entered an era of precision test. Not only stringent
constraints on new physics beyond the SM have been derived, but
also information about some of the more subtle aspects on the low
energy strong interaction responsible for $B$ hadronization and
related hadronic decay amplitudes has been extracted, in
particular for $B$ to two $SU(3)$ octet pesudoscalars
$PP$\cite{rosner1,he1}. In this work we study information about
decay amplitudes and CP asymmetries using
the recently measured branching ratios for\cite{data} $B^+\to
\pi^+ K^0, K^+ \bar K^0$ and $B^0\to K^0\bar K^0$.

There are several reasons why $B^+\to \pi^+ K^0, K^+\bar K^0$ and
$B^0 \to K^0\bar K^0$ decays are interesting to
study\cite{rosner2}. The tree contributions are usually thought to
be small for these decays since they come from the so called
annihilation diagrams and are neglected in many of the previous
studies. If the annihilation contributions are neglected, $\tilde
B(B^+\to \pi^+\ K^0)/\tilde B(B^+\to K^+ \bar K^0)$ (here the
branching ratios are averaged over $B$ and $\bar B$ decays) is
approximately equal to $|V_{ts}/V_{td}|^2$ which can be tested. A
method using $B\to \pi K$ decays to determine the CP violating
phase $\gamma$ in the Cabbibo-Kobayashi-Maskawa (CKM) matrix as
discussed in Ref.\cite{neubert} crucially depends on the
assumption that the annihilation contribution to $B^+\to \pi^+
K^0$ is zero. Therefore it is important to find out whether the
annihilation contributions is zero or not. Using recent
experimental data, we find that the annihilation contributions to
rare B decays may not be negligible after all. We also find that
using a relation in the amplitudes for $B^+\to \pi^+ K^0$ and
$B^+\to K^+ \bar K^0$, with non-zero annihilation contributions
there is a relation for CP asymmetries in $B^+\to \pi^+ K^0$ and
$B^+\to K^+ \bar K^0$. From this relation, we obtain an upper
bound of 5\% for the size of CP asymmetry in $B^+\to \pi^+ K^0$.
The size of CP asymmetry in $B^+\to K^+\bar K^0$ can, however, be
as large as 90\%. In the following we provide details.

In the SM the leading order decay amplitude for the charmless $B$
decays can be decomposed into two terms proportional to
$V_{ub}^*V_{uq}$ and $V_{tb}^*V_{uq}$ respectively. For $B^+ \to
\pi^+ K^0$, and $B^+ \to K^+ \bar K^0$, we have
\begin{eqnarray}
&&A(B^+ \to \pi^+ K^0) = V_{ub}^* V_{us} T_s + V_{tb}^*V_{ts}
P_s,\nonumber\\
&&A(B^+ \to K^+ \bar K^0) = V_{ub}^* V_{ud} T_d + V_{tb}^*V_{td}
P_d.
\end{eqnarray}

In the SU(3) limit, $T_s=T_d $ and $P_s=P_d$. In terms of the
diagram amplitudes discussed in Ref. \cite{rosner3}, one can
parameterize them as $T_{s} = A$ and $P_s = P -P^C_{EW}/3$, where
$P$, $P_{EW}^C$ and $A$ stand for penguin, electroweak penguin and
annihilation amplitudes, respectively. When $SU(3)$ breaking
effects are included, there are modifications. It is expected that
a large fraction of the $SU(3)$ breaking effects may be absorbed
into the relevant meson decay constants. This can be understood
from the PQCD calculation for $B \to PP$ decay
amplitudes\cite{pqcd}. In this approach quarks are viewed as
partons forming the initial and final mesons and are convoluted
with the relevant hadron light-cone wave functions which are
normalized to their decay constants. To the leading order one
would have $f_K T_d = f_\pi T_s$ and $f_K P_d = f_\pi P_s$. We can
then further simplify the above decay amplitudes to
\begin{eqnarray}
&&A(B^+ \to \pi^+ K^0) = P_s \left ( V_{ub}^* V_{us} a  +
V_{tb}^*V_{ts}\right ),\nonumber\\
&&A(B^+ \to K^+ \bar K^0) = {f_K\over f_\pi} P_s \left ( V_{ub}^*
V_{ud} a + V_{tb}^*V_{td}\right ),\label{eq1}
\end{eqnarray}
where $a = T_s/P_s$ is a complex number. We parameterize it
as $a = r e^{i\delta_r}$ for our later discussions.

For the above two processes only small annihilation amplitude $A$
contributes to terms proportional to $V_{ub}V^*_{uq}$, they are usually
neglected. If true, there are several
interesting experimental consequences with one of them being that
CP asymmetries in $B^+ \to \pi^+ K^0$, and $B^+ \to K^+ \bar K^0$
are zero, and also that $R = \tilde B(B^+\to \pi^+ \bar
K^0)/\tilde B(B^+ \to K^+ \bar K^0)$ would give a good
determination for $(f^2_\pi/f^2_K)|V_{ts}/V_{td}|^2$.

Recent experiments at the BaBar and Belle have measured branching
ratios of these decays with\cite{data} $\tilde B(B^+\to \pi^+ K^0)
= (23.1 \pm 1.0)\times 10^{-6}$ and $\tilde B(B^+\to K^+  K^0) =
(1.36^{+0.29}_{-0.27})\times 10^{-6}$. Using these numbers and
$f_K/f_\pi =1.198\pm 0.003^{+0.016}_{-0.005}$\cite{pdg,lat}, we
obtain,
\begin{eqnarray}
{|V_{td}|^2\over |V_{ts}|^2} = 0.041\pm 0.009,
\end{eqnarray}
in accordance with best global fit value for the CKM parameters
which gives\cite{pdg} $|V_{td}|^2/|V_{ts}|^2 = 0.036$.

If the $SU(3)$ correction factor $f_K/f_\pi$ in eq.(\ref{eq1}) is
not included, one would obtain a value $|V_{td}|^2/|V_{ts}|^2$
around 0.059 which is substantially away from the best global fit
value. This supports the expectation that a large fraction of the
$SU(3)$ breaking effects are taken care of by the light meson
decay constants. We note that the central value of
$|V_{td}|^2/|V_{ts}|^2$ determined using $f^2_\pi/f^2_K R$ and
using global fit deviate from each other. Of course it is too
early to say there is a definitive difference due to large errors.

\begin{figure}[htb]
\begin{center}
\includegraphics[width=8cm]{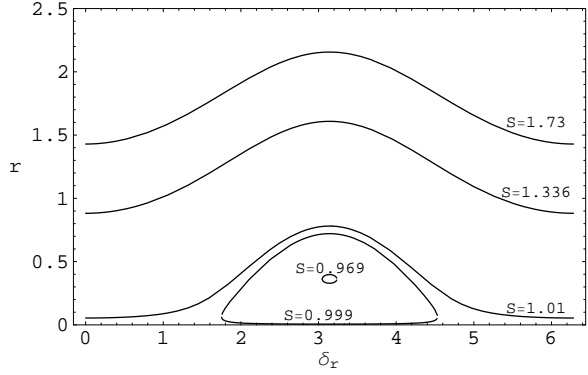}
\end{center}
\caption{ Allowed ranges for $r$ and $\delta$ for different values
of  $ S =(\tau_{B^0}/\tau_{B^+})( \tilde B(B^+\to K^+ \bar
K^0)/\tilde B(B^0\to K^0\bar K^0))$. There is no solution if $S$
is smaller than 0.968.} \label{fig0}
\end{figure}

The available experimental data also allow us to extract more
detailed information about the decay amplitudes. The amplitudes
$T_{d,s}$ need not to be zero. In has been pointed out before that
the data allow non-zero values for $T_{d,s}$\cite{he1}. The
discussion above also does not rule out this possibility. We would
like to point out that there is another indication that the
amplitude $A$ is not zero. This is from a comparison of the
observed branching ratios for $B^+ \to K^+ \bar K^0$ and $B^0\to
K^0\bar K^0$. One can write the amplitudes for these decays as
\begin{eqnarray}
&&A(B^+ \to K^+ \bar K^0) = V_{ub}^*V_{ud} A +V_{tb}^*V_{td}
(P-{1\over 3} P_{EW}^C),\nonumber\\
&&A(B^0 \to K^0 \bar K^0) = V_{tb}^*V_{td}(P-{1\over 3} P_{EW}^C
+P_A).
\end{eqnarray}
Note that one needs to introduce a new amplitude $P_A$ which is
called the penguin-annihilation amplitude, for $B^0\to K^0 \bar
K^0$. Since the Wilson coefficients involved in $P_A$ are smaller
than those in $A$, $P_A$ is expected to be smaller than $A$ in
size. If annihilation amplitudes $A$ and $P_A$ are negligibly
small, one would obtain $S =(\tau_{B^0}/\tau_{B^+})(\tilde
B(B^+\to K^+ \bar K^0)/\tilde B(B^0\to K^0\bar K^0)) = 1$.
However, the measured branching ratios\cite{data} $\tilde B(B^0
\to K^0\bar K^0) = (0.95^{+0.20}_{-0.19})\times 10^{-6}$ and
$\tilde B(B^+\to K^+ \bar K^0)= (1.36^{+0.29}_{-0.27})\times
10^{-6}$ gives $S=1.32\pm 0.39$. The central value of $S$ is
substantially away from 1. This indicates that the annihilation
amplitude $A$ or $P_A$ (or both of them) maybe not zero. To have
some detailed idea about what are the allowed ranges for $r$ and
$\delta_r$, we consider the case with $P_A$ neglected in more
detail. We have
\begin{eqnarray}
S = 1+ r^2 \left | {V_{ub}^*V_{ud}\over V_{tb}^*V_{td}}\right |^2
+ 2 r \cos\delta_r Re\left ({V_{ub}^*V_{ud}\over
V_{tb}^*V_{td}}\right ).
\end{eqnarray}
Using the above we can obtain the allowed ranges for $r$ and
$\delta_r$. In Fig.\ref{fig0} we show $r$ as a function of
$\delta_r$ for several values of $S$.  In obtaining
Fig.\ref{fig0}, we have treated the CKM matrix elements as known
and used the most recent values given by the Particle Data Group
with\cite{pdg} $\lambda = 0.2262$, $A=0.815$ $\overline{\rho} =
0.235$ and $\overline{\eta} = 0.349$ (the corresponding sines of
mixing angles and phase are: $s_{12}=0.2262$, $s_{13}=0.0039$,
$s_{23}=0.0417$ and $\delta_{13}=0.9781$).

There are theoretical calculations\cite{theory, beneke} trying to
estimate the decay amplitudes for $B^+\to \pi^+ K^0$ and $B^+\to
K^+\bar K^0$. One of the popular methods used to estimate such
contributions is the QCDF\cite{beneke}. In this method, the
annihilation amplitudes have end point divergences which are
usually indicated by a quantity $X_A = \int^1_0 dy/y$. In
Ref.\cite{beneke} this is phenomenologically regulated by a cut
off $\Lambda_b$ parameterized as $X_A = (1+ \rho_A
e^{i\phi_A})\ln(m_B/\Lambda_b)$ and $\rho_A < 1$. The resulting
annihilation amplitude $A$ is in general complex. The typical
value from such a calculation gives $r$ of order 10\% with a cut
off $\Lambda_b \sim 0.5$ GeV. If a smaller cut off $\Lambda_b$ is
used, $r$ can be larger. Owing to the uncertainties in treating
the divergences, one cannot give a precise value for
$a=re^{\delta_r}$. Therefore in our later discussions, we will not
use theoretical value for $a$ but use allowed range from data. The
important point is that one should not set $r$ to zero for
precision studies.  We now study some implications with non-zero
annihilation contributions. Before carrying out an numerical
analysis of the allowed ranges for $r$ and $\delta_r$, we study
some implications for CP violation with non-zero values of $r$ and
$\delta_r$. One consequence is that CP violating asymmetries
$A_{CP}(B^+\to \pi^+ K^0, K^+ \bar K^0)$ defined in the following
will not be zero,
\begin{eqnarray}
A_{CP}(B^+\to \pi^+ K^0) &=& {B(B^-\to \pi^- \bar K^0)-B(B^+\to
\pi^+  K^0) \over B(B^- \to \pi^- K^0) + B(B^+\to \pi^+
\bar K^0)}\nonumber\\
 &=& -{2 r \sin\delta_r Im(V_{ub}^*V_{us}
V_{tb}V_{ts}^*)\over |V_{tb}^*V_{ts}|^2 + r^2|V_{ub}^*V_{us}|^2 +
2 r \cos\delta Re(V_{ub}^*V_{us}V_{tb} V_{ts}^*)},\nonumber\\
A_{CP}(B^+\to K^+ \bar K^0) &=& {B(B^-\to K^-  K^0)-B(B^+\to K^+
\bar K^0) \over B(B^- \to K^- K^0) + B(B^+\to K ^+
\bar K^0)}\nonumber\\
 &=& -{2 r \sin\delta_r Im(V_{ub}^*V_{ud}
V_{tb}V_{td}^*)\over |V_{tb}^*V_{td}|^2 + r^2|V_{ub}^*V_{ud}|^2 +
2 r \cos\delta Re(V_{ub}^*V_{ud}V_{tb} V_{td}^*)}.
\end{eqnarray}

It has been pointed out before\cite{he2} that, due to the property
$Im(V_{ub}^*V_{us} V_{tb}V_{ts}^*) = - Im(V_{ub}^*V_{ud}
V_{tb}V_{td}^*)$ of the CKM matrix, there are relations between
rate differences $\Delta(PP) = \Gamma(\bar B \to \bar P\bar P) -
\Gamma(B \to PP)$ for some $B\to PP$ decays. For $A_{CP}(B^+\to
\pi^+ K^0)$ and $A_{CP}(B^+\to K^+ \bar K^0)$, we have
\begin{eqnarray}
A_{CP}(B^+\to \pi^+ K^0) = - {f^2_\pi\over f^2_K} {1\over R}
A_{CP}(B^+ \to K^+ \bar K^0).\label{eq2}
\end{eqnarray}

Since the maximal size for  $A_{CP}(B^+\to K^+\bar K^0)$ can at most be
1, one immediately obtains an upper bound of 5\% for the size
of CP asymmetry in $B^+\to \pi^+ K^0$. The present data $0.009 \pm
0.025$ on $A_{CP}(B^+\to \pi^+ K^0)$ is consistent with the upper
bound. Measurement of CP asymmetry in $B^+\to \pi^+ K^0$ therefore
can test the relation in eq.(\ref{eq2}). Using experimental data for 
$A_{CP}(B^+ \to \pi^+ K^0)$, one would obtain,
$A_{CP}(B^+\to K^+ \bar K^0) \approx - 0.22\pm 0.62$. This is to
be compared with the data $0.12^{+0.17}_{-0.18}$ for
$A_{CP}(B^+\to K^+\bar K^0)$. Naively it seems that there is a 
potential problem since 
the central experimental
data has opposite sign as that  predicted. However, the error
bar is still too large to draw a conclusion. Improved data on
$B^+\to K^+\bar K^0$ can provide more crucial information about
hadronic parameters in $B$ decays.

To find out how large CP asymmetry in $B^+\to K^+ \bar K^0$ can be
and how it is correlated to information from $B^+\to \pi^+ K^0$,
more information is needed. Again we use the experimental data on
the branching ratios and known values for the CKM matrix elements
to constrain the allowed regions. Theoretically, we have
\begin{eqnarray}
R = {\tilde B(B^+\to \pi^+ K^0)\over \tilde B(B^+ \to K^+ \bar
K^0)} ={f^2_\pi\over f^2_K} {|V_{tb}^*V_{ts}|^2 +
r^2|V_{ub}^*V_{us}|^2 + 2 r \cos\delta_r Re(V_{ub}^*V_{us}V_{tb}
V_{ts}^*)\over |V_{tb}^*V_{td}|^2 + r^2|V_{ub}^*V_{ud}|^2 + 2 r
\cos\delta_r Re(V_{ub}^*V_{ud}V_{tb} V_{td}^*)}.
\end{eqnarray}

\begin{figure}[htb]
\begin{center}
\includegraphics[width=8cm]{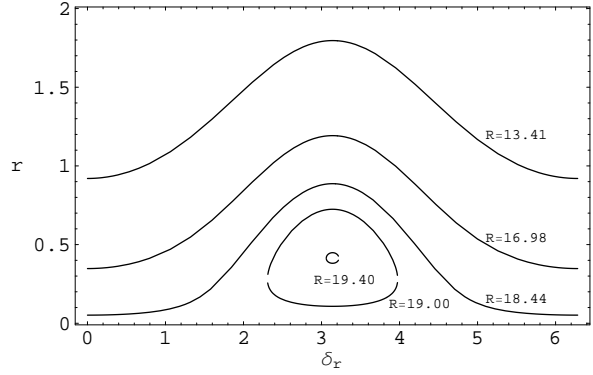}
\end{center}
\caption{ Allowed ranges for $r$ and $\delta$ for different values
of  $ R = \tilde B(B^+\to \pi^+ K^0)/\tilde B(B^+\to K^+\bar
K^0)$.} \label{fig1}
\end{figure}

Comparing the above expression with data, we obtain the allowed
ranges of the parameter space for $r$ and $\cos\delta_r$ in
Fig.\ref{fig1}.  We find that not the entire allowed one-$\sigma$
range of $R = 13.41\sim 20.55$ has solutions for $r$ and
$\delta_r$. With increased $R$ the allowed range becomes smaller
and eventually shrinks to a point at $R = 19.44$. There are no
solutions beyond that point. If future experimental data determine
a $R$ larger than 19.44, it is a strong indication that there is
new physics beyond the SM so that the relation in eq.(\ref{eq1})
is badly broken. Within the one-$\sigma$ allowed range for $R$,
there is a solution for $r=0$ where $R=18.65$. With more precise
determination of $R$, one can have better information about the
size of $r$. At present we see that there is a large allowed range
for $r$ and $\delta_r$. Although large $r$ is unlikely from
theoretical considerations, this possibility is not ruled out yet
at present. We note that the allowed range is more restrictive
than that shown in Fig.\ref{fig0} if both $R$ and $S$ are
restricted to be within their one-$\sigma$ ranges. In the
following discussions we will use the ranges for $r$ and
$\delta_r$ constrained from $R$ instead that from $S$. The reason
is two folds with one of them being that the constraint from $R$
is more stringent, and another being that in obtaining allowed
ranges for $r$ and $\delta_r$ in Fig.\ref{fig0}, $P_A$ has been
neglected. A non-zero $P_A$ can change the ranges.

\begin{figure}[htb]
\begin{center}
\includegraphics[width=8cm]{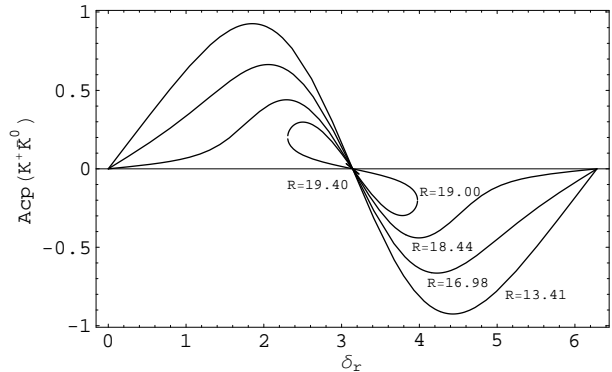}
\end{center}
\caption{ $A_{CP}(B^+\to K^+ \bar K^0)$ as a function of $\delta$
for several allowed values $R$.}\label{fig2}
\end{figure}

With the allowed ranges for $r$ and $\delta_r$ fixed, one can
obtain the allowed range for CP asymmetry in $B^+\to K^+\bar K^0$.
We present the predicted CP asymmetry for $B^+\to K^+ \bar K^0$ in
Fig.\ref{fig2}. CP asymmetry for $B^+\to \pi^+ K^0$ can be easily
obtained from Fig.\ref{fig2} using eq.(\ref{eq2}). We see that the
size of CP asymmetry $A_{CP}(B^+\to K^+\bar K^0)$ becomes larger
as $R$ decreases. At the lower end of the one-$\sigma$ allowed
$R$, the size of $A_{CP}(B^+\to K^+ \bar K^0)$ can be as large as
0.92. Consequently the size of $A_{CP}(B^+\to \pi^+ K^0)$ can
almost reach its upper bound of 0.05. Measurement on CP
asymmetries in these decay modes can provide valuable information
about the hadronic parameters in $B$ decays.

To summarize, in this work we have shown that the recently
measured branching ratios for $B^+\to K^+\bar K^0$ and $B^0 \to
K^0 \bar K^0$ indicate that the annihilation amplitudes in these
decays may be not zero. This observation has important impacts on
CP violation
in $B^+ \to \pi^+ K^0, K^+ \bar K^0$. Using the measured branching
ratios for these decays, we find that there is an absolute bound
of $5\%$ for the size of CP asymmetry in $B^+\to \pi^+ K^0$ from
a relation in the amplitudes of these decays. The size of CP
asymmetry in $B^+ \to K^+\bar K^0$ can, however, be as large as
$90\%$. Future experimental data will test these predictions.

\acknowledgments This work was supported in part by NNSF and NSC.
XGH thanks NCTS for partial support.


\begin{references}

\bibitem{rosner1}
Y.-L. Wu, Y.-F. Zhou and C. Zhuang, hep-ph/0609006;
hep-ph/0606035; Peng Guo, X.-G. He and X.-Q. Li, Int. J. Mod.
Phys. {\bf A21}, 57(2006); C.S. Kim et al., hep-ph/0509015; Y.-L.
Wu and Y.-F. Zhou, Phys. Rev. {\bf D72}, 034037(2995); Phys. Rev.
{\bf D71}, 021701(2005); D. Chang et al., hep-ph/0510328; C.S. S.
Oh and C. Yu, Phys. Rev. {\bf D72}, 074005(2005); A. Buras et al.,
Phys. Rev. Lett. {\bf 92}, 101804(2004); S. Baek et al., Phys.
Rev. {\bf D71}, 057502(2005); X.-G. He and B. McKellar,
hep-ph/0410098; T. Carruthers and B. McKellar, hep-ph/0412202;
Y.-Y. Charng and H.-n. Li, Phys. Rev. {\bf D71}, 014036(2005); S.
Mishima and T. Yoshikawa, Phys. Rev. {\bf D70}, 094024(2994);
Phys. Lett. {\bf B594}, 185(2004); C.-W. Chiang et al., Phys. Rev.
{\bf D70}, 034020(2004); Z.-J. Xiao, C.-D. Lu and L. Gio,
hep-ph/0303070.

\bibitem{he1} A. Buras et al., Eur. Phys. J. {\bf C32}, 45(2003);
H.-K. Fu, X.-G. He and Y.-K. Hsiao, Phys. Rev. {\bf D69},
074002(2004); H.-K. Fu et al., Chin. J. Phys. {\bf 41}, 601(2003);
X.-G. He, et al., Phys. Rev. {\bf D64}, 034002(2001); Y.-F. Zhou
et al., Phys. Rev. {\bf D63}, 054011(2001).

\bibitem{data} BABAR Collaboration, B. Aubert et al., ICHEP 06,
Russia, 26 Jul - 2 Aug. 2006, hep-ex/0608036; Belle Collaboration,
K. Abe et al., hep-ex/0608049;
http://www.slac.stanford.edu/xorg/hfag.

\bibitem{rosner2} M. Gronau and J. Rosner, Phys. Rev. {\bf D58},
113005(1998); D. Atwood and A. Soni, Phys. Rev. {\bf D58},
036005(1998); D. Delepine, J. Gerard, J, Pestieau and J. Weyers,
Phys. Lett. {\bf B429}, 106(1998); A. Falk, A. Lagan, Y. Nir and
A. Petrov, Phys. Rev. {\bf D57}, 4290(1998); R. Fleischer, Phys.
Lett. {\bf B435}, 221(1998);


N.G. Deshpande et al., Phys. Rev. Lett. {\bf 83}, 1100(1999); P.
Zenczykowski, Phys. Rev. {\bf D63}, 014016(2001).

\bibitem{neubert} M. Gronau, D. London and J. Rosner, Phys. Rev. Lett. {\bf 73}, 21(1994);
N.G. Deshpande and X.-G. He, Phys. Rev. Lett. {\bf 74}, 26(1995);
M. Neubert and J. Rosner, Phys. Rev. Lett. {\bf 81}, 5076(1998).

\bibitem{rosner3} M. Gronau et al., Phys. Rev. {\bf D50},
4529(1994); {\bf D52}, 6374(1995).

\bibitem{pqcd} Y.-Y. Keum, H.-N. Li and I.A. Sanda, Phys. Rev.
{\bf D63}, 054008(2001); Phys. Lett. {\bf B504}, 6(2001); C.-D.
Lu, K. Ukai and M.-Z. Yang, Phys. Rev. {\bf D63}, 074009(2001).

\bibitem{pdg} Particle Data Group, J. Phys. {\bf G33}, 1(2006).

\bibitem{lat} MILC Collaboration, C. Bernard et al., Lattice 2005,
Trinity College, Dublin, Ireland, 25-30, Jul 2005 {\bf Pos
LAT2005}"025(2006)[hep-lat/0509137].

\bibitem{theory} C.-D. Lu and Y.-L. Shen and W. Wang, Phys. Rev. {\bf D73}, 034005(2006);
C.-H. Chen and H.-n. Li, Phys. Rev. {\bf D63},
014003(2001).

\bibitem{beneke} M. Bebeke et al.,, Nucl. Phys. {\bf B606}, 245(2001).

\bibitem{he2} N.G. Deshpande and X.-G. He, Phys. Rev. Lett. {\bf 75},
1703(1995);  X.-G. He, Eur. Phys. J. {\bf C9}, 443(1999); M.A.
Dariescu, et. al., Phys. Lett. {\bf B557}, 60(2003).


\end{references}
\end{document}